\documentclass[9pt, conference]{IEEEtran}
\IEEEoverridecommandlockouts

\usepackage{mathtools}
\usepackage{amsfonts}
\usepackage{dsfont}
\usepackage{caption}
\usepackage{subcaption}
\usepackage{lipsum}
\usepackage{tikz}%<<<<<<<<<<<<<<
\usepackage{nicematrix}%<<<<<<<<<<<<<<

\usepackage{cite}
\usepackage{amsmath,amssymb,amsfonts}
\usepackage{algorithmic}
\usepackage{graphicx}
\usepackage{textcomp}
\usepackage{xcolor}

\newcommand{\rulesep}{\unskip\ \vrule\ }

\def\BibTeX{{\rm B\kern-.05em{\sc i\kern-.025em b}\kern-.08em
    T\kern-.1667em\lower.7ex\hbox{E}\kern-.125emX}}
    
\begin{document}

\title{Variable Bitrate Residual Vector Quantization \\ for Audio Coding
}

\author{\IEEEauthorblockN{
    \textit{
    Yunkee Chae$^{1,3}$$^\dagger$
    \thanks{$^\dagger$
        Work done during an internship at Sony AI and partly supported by 
        Institute of Information \& communications Technology Planning \& Evaluation (IITP) grant funded by the Korea government(MSIT) [No. RS-2022-II220320, 2022-0-00320, 40\%], [No. RS-2022-II220641, 50\%],  [No.RS-2021-II211343, Artificial Intelligence Graduate School Program (Seoul National University), 5\%], and [No.RS-2021-II212068, Artificial Intelligence Innovation Hub (Artificial Intelligence Institute, Seoul National University), 5\%]. We used ABCI 2.0 provided by AIST and AIST Solutions.
    },
    Woosung Choi$^{1}$,
    Yuhta Takida$^{1}$,
    Junghyun Koo$^{1}$,
    Yukara Ikemiya$^{1}$,
    % Zhi Zhong$^{2}$
    % Kin Wai Cheuk$^{2}$, 
    % Marco Martinez Ramirez$^{2}$,
    % Kyogu Lee$^{1}$,
    % Wei-Hsiang Liao$^{2}$,
    % Yuki Mitsufuji$^{2}$,
    }}
    \IEEEauthorblockN{
    \textit{
    % Yunkee Chae$^{1}$,
    % Woosung Choi$^{2}$,
    % Yuhta Takida$^{2}$,
    % Junghyun Koo$^{2}$,
    % Yukara Ikemiya$^{2}$,
    Zhi Zhong$^{2}$,
    Kin Wai Cheuk$^{1}$, 
    Marco A. Martínez-Ramírez$^{1}$,
    Kyogu Lee$^{3,4,5}$,
    Wei-Hsiang Liao$^{1}$,
    Yuki Mitsufuji$^{1,2}$
    }\vspace{.4\baselineskip}}
    \IEEEauthorblockA{
    {$^{1}$Sony AI, $^{2}$Sony Group Corporation, Tokyo, Japan} \\
    % {$^{1}$Sony AI, Tokyo, Japan} \;
    % {$^{2}$Sony Group Corporation, Tokyo, Japan} \\    
    % {$^{3}$Interdisciplinary Program in Artificial Intelligence, $^{4}$Department of Intelligence and Information,} \\
    % {$^{4}$Department of Intelligence and Information, Seoul National University} \\
    % {$^{5}$Interdisciplinary Program in Artificial Intelligence, Seoul National University} \\
    % { $^{5}$Artificial Intelligence Institute, Seoul National University} \\
    {$^{3}$IPAI, $^{4}$AIIS, $^{5}$Department of Intelligence and Information, Seoul National University}
    }
}

\maketitle

\begin{abstract}
Recent state-of-the-art neural audio compression models have progressively adopted residual vector quantization (RVQ).
Despite this success, these models employ a fixed number of codebooks per frame, which can be suboptimal in terms of rate-distortion tradeoff, particularly in scenarios with simple input audio, such as silence.
To address this limitation, we propose variable bitrate RVQ (VRVQ) for audio codecs, which allows for more efficient coding by adapting the number of codebooks used per frame.
Furthermore, we propose a gradient estimation method for the non-differentiable masking operation that transforms from the importance map to the binary importance mask, improving model training via a straight-through estimator. 
We demonstrate that the proposed training framework achieves superior results compared to the baseline method and shows further improvement when applied to the current state-of-the-art codec.
Audio samples are available at: https://yoongi43.github.io/VBRRVQ.github.io/
\end{abstract}

\section{Introduction}\label{sec:intro}
An audio codec encodes/decodes the audio, aiming to make the decoded signal perceptually indistinguishable from the original while using the minimum number of codes to conserve the bit budget.
Conventional audio codecs usually rely on expert knowledge of psycho-acoustics and traditional digital signal processing to code and allocate bits to represent the coefficient of time-frequency domain representation such as modified discrete cosine transform (MDCT) \cite{princen1987subband}.
The emergence of deep neural networks (DNNs) has led to the development of neural audio codecs that outperform conventional ones. 
Notably, residual vector quantization (RVQ)-based neural audio codecs \cite{zeghidour2021soundstream, defossez2022highencodec, kumar2024highdac} have demonstrated state-of-the-art performance by producing compact discrete representations at low bitrates. 
These codecs encode raw waveforms into parallel token sequences, subsequently decoded to reconstruct the audio signal from the summations of quantized residual vectors. 

Since a typical RVQ-based neural codec allocates the same number of codebooks across all downsampled time frames, it can be considered a form of constant bitrate (CBR) codec.
This approach can be inefficient for encoding less complex segments, such as silence.
Traditional codecs like AAC \cite{aac} tackle this inefficiency by supporting variable bitrate (VBR), which allocates more bits to complex audio segments and fewer bits to simpler ones.
Recently, SlowAE \cite{dieleman2021variable} introduced VBR discrete representation learning by modeling the event-based representations using run-length transformers along the time dimension. 
Additionally, a VBR codec using DNN-based vector quantization is proposed in    \cite{li2021variable}, employing a code allocation scoring function to determine the number of vectors allocated for each audio window. 
However, introducing VBR to an RVQ-based neural audio codec has not been explored yet. 

Meanwhile, similar efforts have been made for VBR in image compression.
For instance, \cite{baldassarre2023variable} utilized the self-attention map of a self-supervised DINO model \cite{caron2021emerging} as a proxy for capacity to be assigned, leveraging the model's capability to capture semantic saliency. 
They allocated a variable number of bits to each pixel, applying a bit allocation map to their product quantizer \cite{jegou2010product}.
On the other hand, some works have introduced \textit{importance map} \cite{li2018learning, mentzer2018conditional, li2020learning, lee2022selective} to determine how many channels per pixel in the quantized features should be used.
These importance maps are trained jointly with the compression model, enabling effective bitrate allocation for each pixel.

Inspired by these approaches, we propose a VBR RVQ (VRVQ) model, which allocates a different number of bits to each downsampled time frame in the latent space based on an importance map.
By being jointly trained with rate-distortion loss optimization, the importance map learns to determine the optimal number of codebooks per frame, leading to a more effective rate-distortion curve compared to previous RVQ-based codecs.
Consequently, our contributions are as follows:\\
\textbf{Variable Bitrate RVQ}:
    We are the first to achieve variable bitrate (VBR) coding to RVQ-VAE-based audio codecs. 
    While many audio codecs utilize VBR, no prior work has attempted to incorporate variable bitrate into RVQ, which is the foundation of current state-of-the-art audio codec models.\\
\textbf{Gradient Estimation for Binarized Variables}:
    We adapt the importance map, originally used in image compression models, for audio coding. 
    The challenge lies in the non-differentiable operations involved in generating binary masks. 
    Previous approaches employed identical backward passes, but these methods proved suboptimal in our context. 
    To address this, we propose a more efficient gradient estimation during binarization, ensuring more effective gradient flow during training.

\begin{figure}
    \centering
    \includegraphics[width=\linewidth,trim=1 1 1 1,clip]{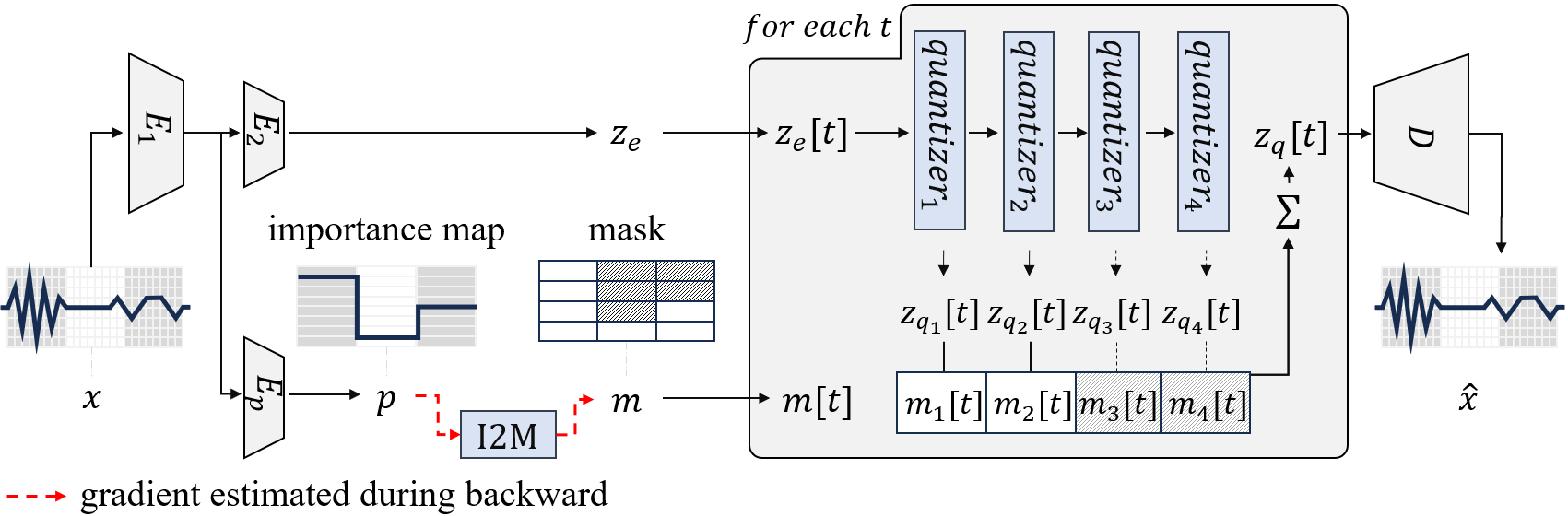}
    \caption{Overall proposed framework of variable bitrate RVQ.}
    \label{fig:overall framework}
    \vspace{-5mm}
\end{figure}

\section{Background}\label{sec:background}
This section summarizes RVQ and structured dropout, used in several state-of-the-art CBR RVQ codecs \cite{zeghidour2021soundstream, defossez2022highencodec, kumar2024highdac}.

\subsection{Residual Vector Quantization}
\label{subsec:rvq}
Let $z_e=E(x) \in \mathbb{R}^{D\times T}$ represent the output of the encoder $E$ for the input audio $x$, where $T$ and $D$ denotes downsampled timeframes and the  dimension of the latent representation, respectively.
We denote the $i$-th frame at the time $t$ of $z_e$ as $z_e[t]$.
An RVQ with $N_q$ quantizers takes as input $z_e[t]$ and computes $z_q[t]$ as follows: 
\begin{equation}
    z_q[t]=\sum_{i=1}^{N_q}Q_i(r_i[t])
    ,
\end{equation}
where $Q_i$ be the $i$-th quantizer and $r_i$ is defined as $r_i[t]=z_e[t]-\sum_{j=1}^{i-1}Q_j(r_j[t])$, with the initial residual given by $r_1[t]=z_e[t]$. The resulting $z_q$ is then fed to the decoder $D$ to output the audio estimation $\hat{x}$.

\subsection{Structured Codebook Dropout}
Given a target bitrate, a separate model can be trained with the corresponding $N_q$. However, instead of this separate training, structured dropout \cite{zeghidour2021soundstream, srivastava2014dropout} can be used to enable a single RVQ model to support multiple target bitrates. During training, rather than utilizing all the codebooks, only the first $n_q$ quantizers are used (i.e., $z_q[t]=\sum_{i=1}^{n_q}Q_i(r_i[t])$), where $n_q$ is randomly sampled from $\{1, \cdots, N_q\}$ for each batch (or batch item \cite{kumar2024highdac}). During inference, we can select the desired $n_q$ for the target bitrate.
We can rewrite the structured codebook dropout using masking notation as follows:
\begin{equation}    z_q[t]=\sum_{i=1}^{N_q}\mathds{1}_{i\leq n_q}\cdot Q_i(r_i[t]),
\label{eq:structured_dropout}
\end{equation}
where $\mathds{1}$ denotes the indicator function.
A neural codec trained with this trick remains a CBR codec, as it uses the same number of codebooks across all frames once a target bitrate is set.

\vspace{5pt}
\section{Variable Bitrate RVQ}\label{sec:vrvq}

We propose a VRVQ by introducing the \textit{importance map}. First, we outline the bit allocation method based on the importance map (\ref{subsec:impmap}). Next, we propose a gradient estimation method (\ref{subsec:gradient_estimation}) to jointly train the importance map within an encoder-decoder framework. Finally, we describe the joint training method (\ref{subsec:training}) in detail.

\subsection{Importance Map-based Bit Allocation}\label{subsec:impmap}

%previously
As shown in Fig. \ref{fig:overall framework}, the proposed bit allocation at time frame index $t$ is controlled by the binary mask $m[t]\in \{0,1\}^{N_q}$, where $m_i[t]$ determines the usage of the $i$-th codebook at $t$.
Note that this mask is time-varying over $t$ unlike the indicator function in Eq. \eqref{eq:structured_dropout}.
We design $m[t]$ to decrease monotonically over $i$ (i.e., $m_i[t] \geq m_j[t]$ for $i < j$) for the compression efficiency.
For a desirable mask generation, we adopt the \textit{importance map}, initially proposed for image compression \cite{li2018learning, li2020learning, lee2022selective, mentzer2018conditional}.
We compute the importance map from an intermediate feature map of the encoder $E$ and generate a mask based on the importance map for the bit allocation.

We start by decomposing $E$ into two components $E=E_2 \circ E_1$ to obtain the intermediate feature map: a shared part $E_1$ and an encoding-specific part $E_2$, borrowing notations from \cite{li2020learning}.
Given an audio $x$, a learnable subnetwork $E_p$ takes $E_1(x)$ as input, and outputs an
importance map $p = E_p (E_1(x)) \in (0, 1)^T$. 
Then, for each time frame index $t$, we generate  $m[t]\in\{0,1\}^{N_q}$ based on $p[t]$ with the following \textit{importance map-to-mask} function $\text{I2M}:(0,1)\rightarrow\{0,1\}^{N_q}$:
\begin{equation}
    m[t]=\text{I2M}(p[t]):=[H^0(S(p[t])),\cdots,H^{N_q-1}(S(p[t]))]^\mathsf{T},
    \label{eq:i2m}
\end{equation}
where $S(p)=N_q\cdot p$ is a scaling function and each $H^k(s)$ is defined as a Heaviside step function as follows:
\begin{align}
H^k(s) := 
\begin{dcases*}
    1 & if $k \leq s$, \\
    0 & if $k > s$
\end{dcases*}.
\label{eq:heaviside}
\end{align}

The obtained mask $m[t]$ is then used to calculate the quantized vector $z_q$ by replacing the indicator function in Eq. \eqref{eq:structured_dropout} as follows:
\begin{equation}    
z_q[t]=\sum_{i=1}^{N_q}m_i[t]\cdot Q_i(r_i[t]).
\label{eq:masked_quantization}
\end{equation}

\subsection{Gradient Estimation for Binarized Variables}
\label{subsec:gradient_estimation}
\subsubsection{Baseline}
Since $H^k$ is a non-continuous function,  gradient estimation for backpropagation is required to jointly train the importance map within the encoding-decoding framework.
In the image domain \cite{li2018learning, mentzer2018conditional},  the saturated identity function $f^k_I$ is employed as a surrogate for $H^k$ during backpropagation, defined as follows, and serves as the baseline function for our work:
\begin{equation}
f^k_I(s):=\max(\min(s-k,1),0)    
\end{equation}

However, this function presents an issue where the gradient does not flow through large regions due to the $\max$ and $\min$ operations (note that non-zero gradient can exist for only a single $k \in \{0, ..., N_q-1\}$ in the $\text{I2M}$ function). 
This problem makes the model suboptimal and degrades the performance of VRVQ, as demonstrated in our experiments.
\subsubsection{Smooth approximation}
\begin{figure}
    \centering
    \includegraphics[width=\linewidth]{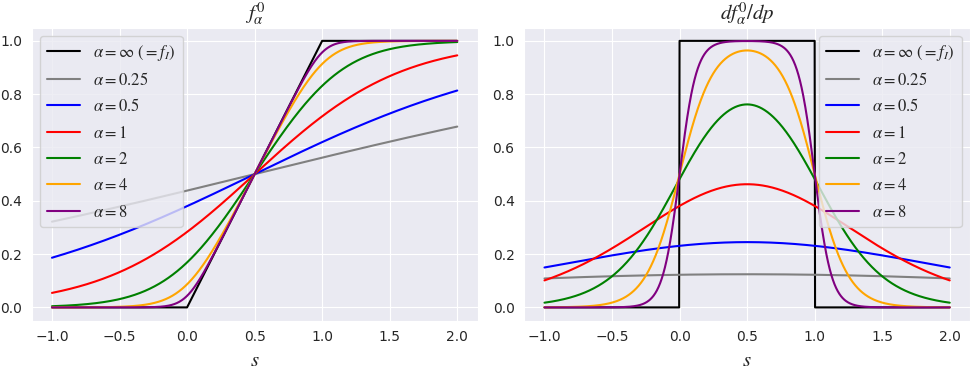}
    \caption{Gradient estimator $f_\alpha^k$ when $k=0$}
    \label{fig:logcosh}
    \vspace{-5mm}
\end{figure}

To address this, we propose a smooth surrogate function that relaxes the $f^k_I$, allowing the gradient to be computed across all ranges, which is defined as follows:
\begin{equation}
    f_\alpha^k(s):=\frac{1}{2\alpha}\log\left(\frac{\cosh(\alpha(s-k))}{\cosh(\alpha(-s+k+1))}\right) + \frac{1}{2}
\end{equation}
This surrogate function is parameterized by $\alpha\in\mathbb{R}_{>0}$, which serves as a hyperparameter. 
In the extreme case where alpha approaches infinity, this function approaches $f^k_I$  for all $s\in\mathbb{R}$ (i.e., $f_\infty^k(x):=\lim_{\alpha\rightarrow\infty}f_\alpha^k(s)=f^k_I(s)$) 
as shown in Fig. \ref{fig:logcosh}.

By substituting the $H^k$ in $\text{I2M}$ with $f_\alpha^k$, we define $\text{I2M}_\text{soft}^\alpha$, the surrogate function of $\text{I2M}$, and then apply the straight-through estimator for backpropagation as follows:
\begin{equation}
    \text{I2M}_\text{soft}^\alpha(p[t]):=[f_\alpha^0(S(p[t])),\cdots,f_\alpha^{N_q-1}(S(p[t]))]
    \label{eq:i2msoft}
\end{equation}
\vspace{-3mm}
\begin{equation}
    p[t]\mapsto \text{I2M}_\text{soft}^\alpha(p[t]) + \text{sg}\left(\text{I2M}(p[t])-\text{I2M}_\text{soft}^\alpha(p[t])\right),
\end{equation}
where $\text{sg}$ denotes the stop-gradient operation.

\begin{figure*}[!t]
    \centering   \includegraphics[width=\linewidth]{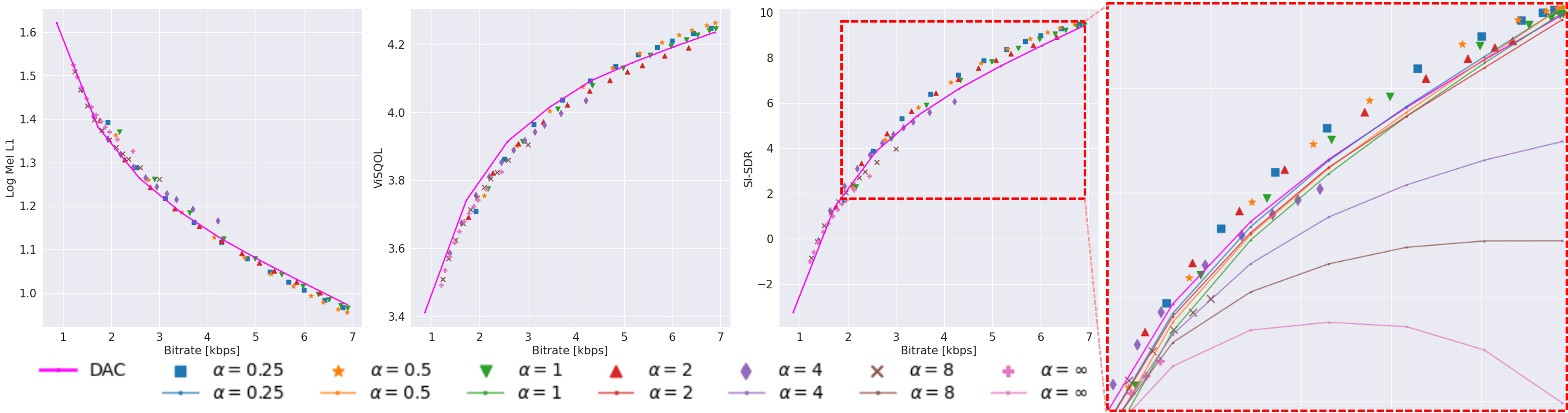}
    % \caption{Caption}
    \caption{The results of VRVQ across different $\alpha$.
    The points marked with various markers represent the results of inference at different scaling factor, $l$=4, 6,  8, 10, 12, 14, 16, 18, 20, 24, and 32, in VBR mode.
    In the rightmost plot, we display solid lines representing the results of inference in CBR mode for each model.
    %% level_list = [4,6,  8, 10, 12, 14, 16, 18, 20, 24, 32]
    % \textit{DAC ckpt} results were measured using the publicly available DAC checkpoint, with up to 8 codebooks utilized during the evaluation.
    }
    \label{fig:results_alpha}
    \vspace{-5mm}
\end{figure*}

\subsection{Training}
\label{subsec:training}
To train the compression model and the importance map subnetwork, we define the rate loss $\mathcal{L}_R$ as follows:
\begin{equation}
     \mathcal{L}_R=\frac{1}{T}\sum_{t=1}^T E_p(E_1(x))[t]
\end{equation}
We basically follow the training scheme of \cite{kumar2024highdac}, such as GAN training. 
During the training phase, we optimize the rate-distortion (R-D) tradeoff loss $\mathcal{L}_{D}+\lambda\mathcal{L}_R$, where $\mathcal{L}_{D}$ represents the training objective proposed in \cite{kumar2024highdac}, and $\mathcal{L}_R$ denotes the proposed rate loss.
The balance between these two components is controlled by the tradeoff parameter $\lambda$, which serves as a hyperparameter.

The downside of the importance map is that once the importance map is trained, it consistently generates fixed output $p$ for the identical input audio $x$, which restricts the flexibility of the model in terms of rate control. 
Inspired by random sampling approach used in structured dropout \cite{srivastava2014dropout, zeghidour2021soundstream}, we address this problem by sampling the scaling factor $l$ from the continuous uniform distribution $U([L_\text{min}, L_\text{max}])$, where $L_
\text{max}>N_q$, replacing the scaling function $S(p)$ in Eqs. \eqref{eq:i2m} and \eqref{eq:i2msoft} with $S_l(p)=l\cdot p$ during training.
This approach enables a single model to handle multiple bitrates in VBR mode.

In this study, we opt not to train a separate entropy model or employ entropy coders such as arithmetic coding. 
Instead, we calculate the bitrate based on the number of codebooks used for discrete representation of the audio on frame-by-frame basis, allowing for potential future enhancements through the incorporation of entropy modeling.

\section{Experiments}
\label{sec:experiments}
\subsection{Model Architecture}\label{subsec:model_arch}
In our experiments, we implement VRVQ audio codecs based on DAC \cite{kumar2024highdac}, which itself builds upon the improved RVQGAN architecture. 
The improved RVQGAN is derived from the Improved VQGAN \cite{yu2021vector}, which employs factorized codes and L2-normalized codebook lookups. 
Specifically, the codebook has an 8-dimensional space with a size of 1024, corresponding to 10 bits per codebook.
The downsampling rate of the time frame is 512, resulting in a frame rate of 86 Hz in the latent space.
In our experiment, we set the maximum number of codebooks $N_q$, to 8.
 
The importance map subnetwork is designed as a simple neural network composed of 1D convolutional layers with weight normalization \cite{salimans2016weight} and Snake activation functions \cite{ziyin2020neural}.
It takes the intermediate feature map $E_1(x)$, which has a channel dimension of 1024, extracted from the layer immediately preceding the encoder's final convolutional block (i.e., $E_2$).
It processes the input feature map through five sequential Conv1D blocks with kernel sizes of [5, 3, 3, 3, 1], progressively reducing the channel dimensions  from 1024 to 512, 128, 32, 8, and finally 1.
The output is then passed through a sigmoid activation function to produce the importance map.

\subsection{Experimental Setup}
\label{subsec:exp_setup}

We set the weight of the rate loss $\lambda=2$, and kept it fixed for all experiments. 
% The bitrate is controlled entirely by choosing the level $l$, with $L_\text{min}=1$ and $L_\text{max}=48$.
During training, we randomly sample the scaling factor $l$ between $L_\text{min}=1$ and $L_\text{max}=48$. 
Note that $l$ can be any positive real number  between $L_\text{min}$ and $L_\text{max}$ for inference.
Every model was trained with a batch size of 32 for 300k iterations, using audio segments of 0.38 seconds in duration.

For evaluation, we measure the ViSQOL \cite{chinen2020visqol}, as well as the SI-SDR \cite{le2019sdr}. 
ViSQOL is a perceptual quality metric ranging from 1 to 5, designed to estimate a mean opinion score. 
We also report the sum of L1 losses of log mel spectrograms, using window sizes of [32, 64, 128, 256, 512, 1024, 2048] with a hop length of 1/4 of each window size, following the same configuration as in \cite{kumar2024highdac}.

\subsection{Dataset}\label{subsec:dataset}

We train our model using a dataset similar to that used in the original DAC model \cite{kumar2024highdac}. 
For the speech dataset, we utilize DAPS \cite{mysore2014can_DAPS}, Common Voice \cite{ardila2019common_commonvoice}, VCTK \cite{veaux2017cstr_VCTK}, and clean speech segments from DNS Challenge 4 \cite{dubey2022icassp2022deepnoise_DNS4}.
For the music dataset, we use MUSDB18 \cite{rafii2017musdb18} and Jamendo \cite{bogdanov2019mtg_jamendo}. 
Additionally, for general sounds, we incorporate AudioSet \cite{gemmeke2017audio_audioset}.
All audio is resampled to 44.1 kHz, and the training samples were normalized to -16 dB LUFS with phase shift augmentation applied.

For evaluation, we use the F10 and M10 speakers from the DAPS dataset, along with the test sets from MUSDB18 and AudioSet. 
We randomly extract 100 10-second segments from each domain, resulting in a total of 300 samples for evaluation. 
The audio is not normalized during the evaluation phase.

\begin{figure*}[!t]
    \centering
    \begin{subfigure}[b]{0.74\columnwidth}
        \centering
        \includegraphics[width=\linewidth]{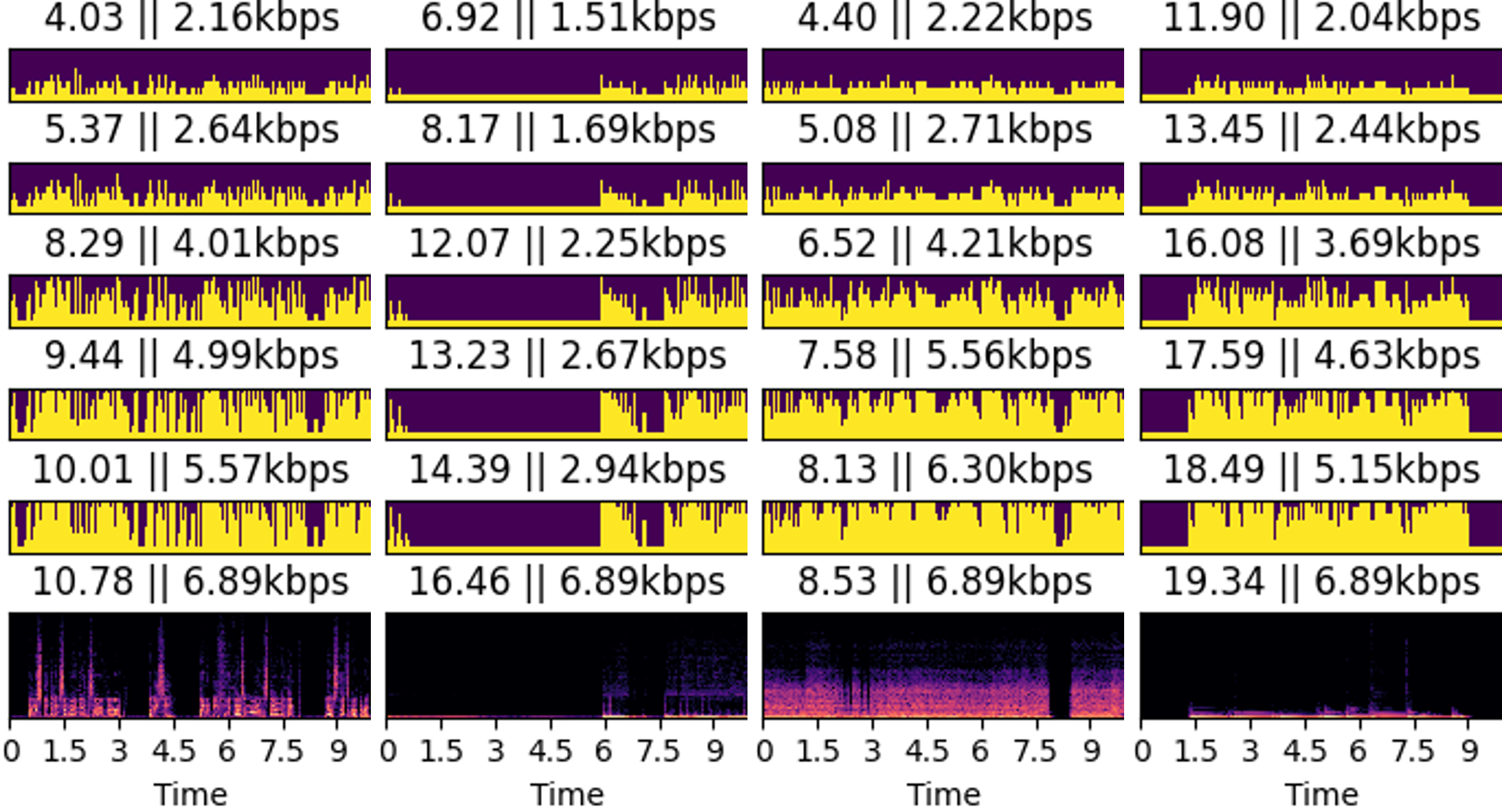}
        % \vspace{-4mm}
        \caption{Codebook usage based on importance map with varying $l$ for four audio samples. The bottom row shows the spectrogram of the input audio.}
        \label{subfig:impmap}
    \end{subfigure}
    \rulesep
    \begin{subfigure}[b]{1.26\columnwidth}
        \centering
        \includegraphics[width=\linewidth]{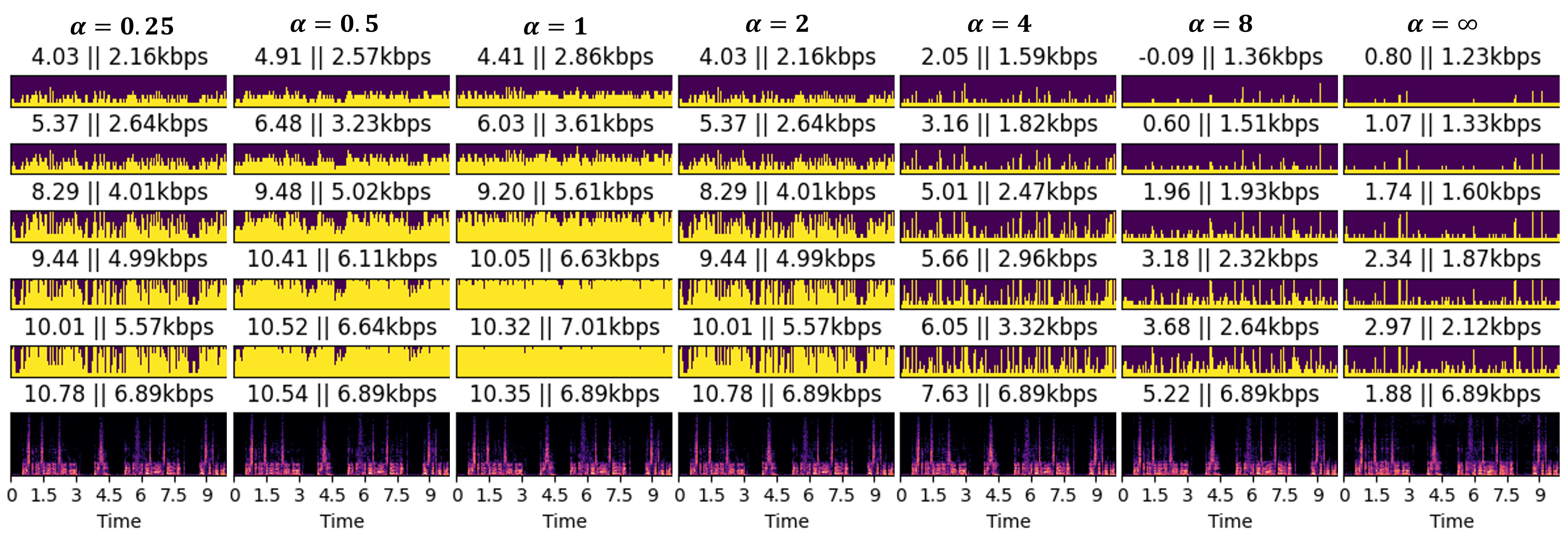}
        % \vspace{-4mm}
        \caption{Codebook useage based on importance map for the same sample with varying $\alpha$ values. 
        The bottom row shows the spectrogram of the reconstructed audio with the full number of codebooks.}
        \label{subfig:alpha_impmap}
    \end{subfigure}
    % \vspace{-1mm}
    \caption{Visualization of the codebook usages. From the first row, the scaling factor $l$ are set to 6, 8, 14, 20, and 26.
    In each plot, the bitrate of the mask and the corresponding SI-SDR are also noted. For the bottom row, the bitrate of the sample is reported when it is inferred with the full number of codebooks in all frames (i.e., 8 codebooks in CBR mode), and thus, the bitrate is calculated without transmission cost.
    }
    \label{fig:imp_map_full_sample}
    \vspace{-2mm}
\end{figure*}
% \vspace{-10mm}

\begin{figure}
    \centering
    \begin{subfigure}[b]{.49\columnwidth}
        \centering
        \includegraphics[width=\linewidth]{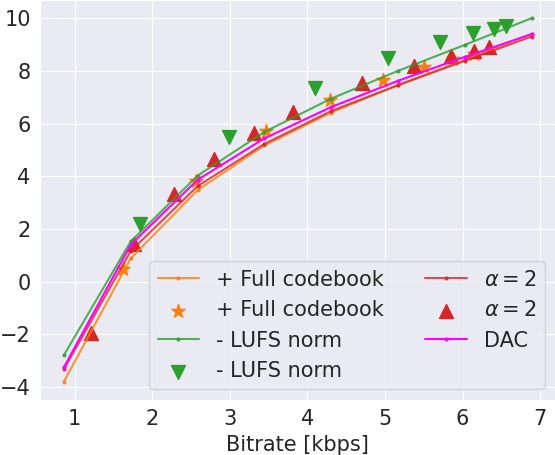}
        % \vspace{-2mm}
        \caption{$N_q=8$}
        \label{subfig:ablation}
    \end{subfigure}
    \begin{subfigure}[b]{.48\columnwidth}
        \centering
        \includegraphics[width=\linewidth]{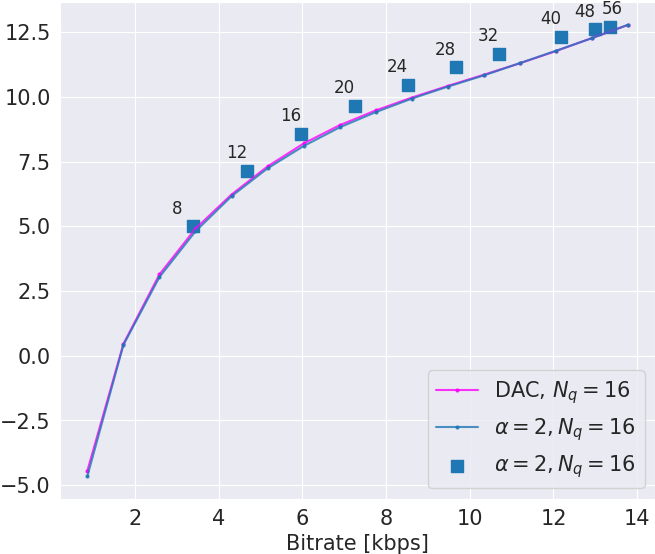}
        % \vspace{-2mm}
        \caption{$N_q=16$}
        \label{subfig:ablation_16}
    \end{subfigure}
    % \vspace{-1mm}
    \caption{SI-SDR results of ablation studies with $N_q=8$ and 16. The scaling factors are set to 4, 8, 12, 16, 20, and 32 in (a), while in (b), they are set to 8, 12, 16, 20, 24, 28, 32, 40, 48, and 56}
    \label{fig:ablations}
    \vspace{-2mm}
\end{figure}

\subsection{Result 1: Comparison on the gradient estimation methods}
Fig. \ref{fig:results_alpha} summarizes the effect of the surrogate function $f_\alpha$ on the performance.
For comparison, we also report the results of CBR RVQ with 8 codebooks trained from scratch for the same number of iterations with structured codebook dropout with a dropout rate of 0.5.
For each model, we plot the rate-distortion curve for each metric.
For our VBR models, we compute the average bitrate over the test dataset as well as the mean of each metric for the scaling factor $l$ in the range of [$L_{\text{min}}$, $L_{\text{max}}$].
% while $l$ increases from $L_{\text{min}}$ to $L_{\text{max}}$.
We also include the additional transmission costs (i.e., $\log_2(N_q)=3$) for the number of codebooks used per frame into the bitrate calculation which amounts to approximately 0.258 kbps. 
Note that our VRVQ models can also be evaluated in a fixed-rate RVQ setup, i.e., in CBR mode not using an importance map. In this case, we do not calculate the transmission cost. 

It is observable that the models trained with the proposed smooth surrogate functions outperform the model trained with the baseline function 
$f_I$.
As $\alpha$ increases, the performance approaches that of $f_I$. 
This observation aligns with the fact that $f_\alpha$ converges to $f_I$.
Our models, trained with $\alpha<2$, show better performance than the DAC trained for the same number of iterations under some conditions.
Notably, for SI-SDR, our models perform significantly better than DAC.
For the log mel loss and ViSQOL, our models outperform the DAC when $\alpha$ is 1 or lower and the bitrate is above 3-4 kbps. 

However, across all metrics, we observe that the benefit of using VBR diminishes at lower bitrates. 
This is likely due to the overhead of additional transmission costs, which become relatively larger for lower bitrates.
The performance at lower bitrates improves for a larger $\alpha$.
This occurs because, as shown in Fig. \ref{subfig:alpha_impmap}, increasing $\alpha$ limits the model's ability to generate the importance map in various ranges, leading it to predominantly sample lower bitrates.
This results in improved performance within that bitrate range.
However, reduced diversity in the importance map results in significant underperformance at higher bitrates. 
It is also notable in Fig. \ref{fig:results_alpha} and \ref{subfig:alpha_impmap} that when $\alpha=\infty$, the SI-SDR of the sample using the full codebook is worse than in cases with lower bitrates.
This is because the importance map during training rarely utilized all codebooks, leading to poorer performance in the full codebook scenario and failing to optimize the rate-distortion tradeoff.

\subsection{Result 2: Ablation Study}
We conduct additional experiments to assess the impact of various configurations. 
First, we investigate whether allocating the full number of codebooks for a subset of samples in each mini-batch during training, as proposed in \cite{kumar2024highdac}, could improve rate-distortion tradeoff, rather than solely relying on importance map-based allocation.
Thus, we conduct ablation studies with $\alpha=2$, where the performance is slightly worse than smaller $\alpha$ values at higher bitrates.
Specifically, for 25\% of the samples in each mini-batch, we use the full set of codebooks regardless of the importance map. 
Additionally, we investigate the effect of removing LUFS normalization preprocessing during training.
We also experiment with increasing the number of codebooks to 16 to assess the scalability of the proposed method with respect to the maximum number of the codebooks.
The corresponding increase in additional transmission cost for VBR is 0.344 kbps. 
For consistency, a DAC model is also trained from scratch using the same configuration.
All ablation studies adhere to the settings outlined in Section \ref{subsec:exp_setup}.
Fig. \ref{fig:ablations} shows how SI-SDR of each model evolves as the bitrate increases.

For models with $N_q=8$, we observed that using the full  codebooks for a subset of samples does not provide a significant advantage.
This is likely because, when $\alpha=2$, the model have observed a sufficiently wide range of the codebook usage cases during training, simply by varying the scaling factor $l$.
Instead, it results in the performance degradation at lower bitrates.
On the other hand, the models trained without LUFS normalization shows a significant performance improvement.
This suggests that the importance subnetwork may have generated a higher-quality importance map by learning the energy distribution of frames.

For $N_q=16$, VRVQ shows better performance than DAC as in the case of $N_q=8$.
In particular, the CBR performance of VRVQ almost matched that of DAC. 
This shows the scalability that our method can be applied to higher bitrates without impairing the performance of existing CBR codecs.

\section{Conclusion}
\label{sec:conclusion}
In this paper, we introduced a VBR RVQ for audio coding, leveraging an importance map. 
While developing the framework, we addressed key challenges such as managing the non-continuous function that converts the importance map into a binary mask by introducing the surrogate functions for gradient estimation, demonstrating that our approach provides better gradient flow compared to the function previously used in the image domain. 
As a result, we achieved efficient bit allocation within the RVQ-VAE framework and demonstrated the effectiveness of our models through rate-distortion curves across various metrics. 
For future work, we plan to explore different model architectures that are better suited for our VRVQ approach, beyond the existing RVQ-based codecs.

\bibliographystyle{IEEEtran}
\bibliography{refs}

\end{document}